\documentclass[nohyperref]{article}

\usepackage{microtype}
\usepackage{graphicx}
\usepackage{subfigure}
\usepackage{booktabs} % for professional tables

\usepackage{hyperref}

\usepackage[accepted]{icml2022}

\usepackage{amsmath}
\usepackage{amssymb}
\usepackage{mathtools}
\usepackage{amsthm}

\usepackage[capitalize,noabbrev]{cleveref}

\theoremstyle{plain}

\theoremstyle{definition}

\theoremstyle{remark}

\newcommand{\mendis}{\texttt{Mendis}}

\usepackage[textsize=tiny]{todonotes}

\icmltitlerunning{Unsupervised Learning for Stellar Spectra with Deep Normalizing Flows}

\begin{document}

\twocolumn[
\icmltitle{Unsupervised Learning for Stellar Spectra with Deep Normalizing Flows}

% It is OKAY to include author information, even for blind
% submissions: the style file will automatically remove it for you
% unless you've provided the [accepted] option to the icml2022
% package.

% List of affiliations: The first argument should be a (short)
% identifier you will use later to specify author affiliations
% Academic affiliations should list Department, University, City, Region, Country
% Industry affiliations should list Company, City, Region, Country

% You can specify symbols, otherwise they are numbered in order.
% Ideally, you should not use this facility. Affiliations will be numbered
% in order of appearance and this is the preferred way.
\icmlsetsymbol{equal}{*}

\begin{icmlauthorlist}
\icmlauthor{Ioana Ciuc\u{a}}{equal,rsaa,soco}
\icmlauthor{Yuan-Sen Ting}{equal,rsaa,soco}
\end{icmlauthorlist}

\icmlaffiliation{rsaa}{Research School of Astronomy \& Astrophysics, Australian National University, Cotter Rd., Weston, ACT 2611, Australia}
\icmlaffiliation{soco}{School of Computing, Australian National University, Acton, ACT 2601, Australia}

\icmlcorrespondingauthor{Ioana Ciuca}{Ioana.ciuca@anu.edu.au}
\icmlcorrespondingauthor{Yuan-Sen Ting}{yuan-sen.ting@anu.edu.au}

% You may provide any keywords that you
% find helpful for describing your paper; these are used to populate
% the "keywords" metadata in the PDF but will not be shown in the document
\icmlkeywords{Machine Learning, ICML}

\vskip 0.3in
]
% \printAffiliationsAndNotice{}
% this must go after the closing bracket ] following \twocolumn[ ...

% This command actually creates the footnote in the first column
% listing the affiliations and the copyright notice.
% The command takes one argument, which is text to display at the start of the footnote.
% The \icmlEqualContribution command is standard text for equal contribution.
% Remove it (just {}) if you do not need this facility.
%\printAffiliationsAndNotice{}  % leave blank if no need to mention equal contribution
\printAffiliationsAndNotice{\icmlEqualContribution} % otherwise use the standard text.
%\printAffiliationsAndNotice{}

\begin{abstract}
Stellar spectra encode detailed information about the stars. However, most machine learning approaches in stellar spectroscopy focus on supervised learning. We introduce \mendis{}, an unsupervised learning method, which adopts normalizing flows consisting of Neural Spline Flows and GLOW to describe the complex distribution of spectral space. A key advantage of \mendis{} is that we can describe the conditional distribution of spectra, conditioning on stellar parameters, to unveil the underlying structures of the spectra further. In particular, our study demonstrates that \mendis{} can robustly capture the pixel correlations in the spectra leading to the possibility of detecting unknown atomic transitions from stellar spectra. The probabilistic nature of \mendis{} also enables a rigorous determination of outliers in extensive spectroscopic surveys without the need to measure elemental abundances through existing analysis pipelines beforehand.
\end{abstract}

\section{Introduction}\label{submission}

As probes of stellar nucleosynthesis, stellar spectra provide an unparalleled testbed for atomic physics in the era of high-resolution stellar spectra from surveys such as GALAH \citep{buder_galah_dr3} and APOGEE \citep{maj_2017}. They offer a window into identifying new atomic lines, which may otherwise evade the notoriously difficult ab initio quantum mechanics calculations \citep{Has_2017, Cunha_2017}. Stellar spectra also reveal the fundamental properties of stars, including their stellar parameters (temperature, surface gravity, and global metallicity) and their chemical composition, which holds the key to unravelling the evolutionary history of our Milky Way galaxy.

Our ever-expanding power to collect spectra has called for more advanced machine learning methods to analyze them. However, most have focused on supervised learning to solve a label determination problem when analyzing spectroscopic data with machine learning \citep{Ness2015, Ting2019}, which is somewhat limiting. On the one hand, there are only a handful of stars with high-fidelity stellar labels (stellar parameters and elemental abundances) \citep{Jofre2014}. On the other hand, due to its inherent complexity, stellar spectral models elude direct physical modelling, even as more advanced theoretical prescriptions become available \citep{Nordlander2017}. This dilemma has often led to perennial debates regarding the pros and cons of applying data-driven or model-driven analysis when dealing with stellar label determination. Additionally, label determination through supervised learning projects the data onto some predefined known label space, significantly impeding the search for exciting unknown outliers.

These challenges pose the question: can we learn from the spectra themselves without first reducing them to stellar labels. Unfortunately, unsupervised learning has not seen the same surge in applications for stellar spectroscopy. Pioneering works of \citet{jones_2018} and \citet{Mijolla_2021} employed methods such as Principal Component Analysis (PCA) and Variational Auto-Enconders (VAE) to extract a representation of stellar spectra. However, these methods are limited in a few key areas. For example, the PCA framework lacks the flexibility to model the data, such as modeling conditional distributions, which are of paramount importance, as we will demonstrate in this study. The posterior embedding distribution learned in a VAE often deviates from the prior distribution and is not able to provide exact likelihood evaluation.

In this study, we explore normalising flows, a class of deep generative models, to bridge this gap. We will show that this more principled way of describing the statistical distribution of stellar spectra through unsupervised learning can lead to new opportunities, including correlation identification and out-of-distribution detection.
\begin{figure*}[ht!]
    \centering
     \includegraphics[width=0.8\textwidth]{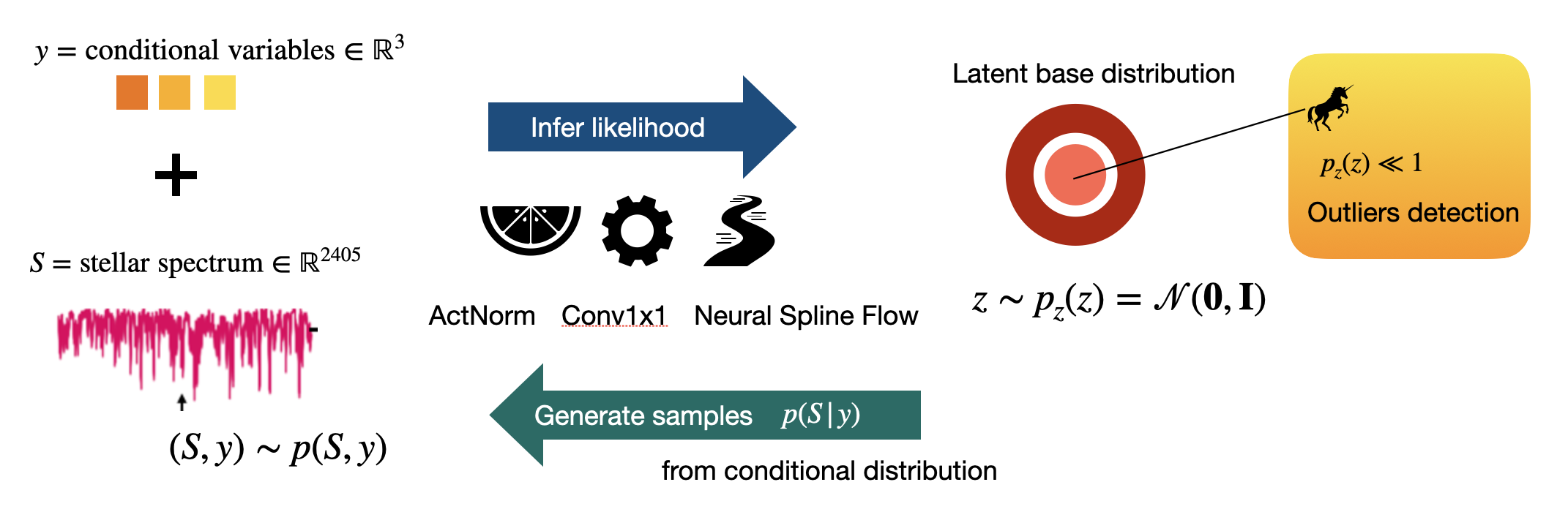}
     \vskip -0.15in
     \caption{Deep normalizing flows for stellar spectra. Our model (\mendis{}) turns an ensemble of training spectra into a Gaussian base distribution from which we can perform exact likelihood evaluation and search for out-of-distribution outliers. As normalizing flows are invertible, they allow us to generate sample from conditional distribution and evaluate all moments and correlations from the distribution.}
     \vskip -0.1in
     \label{fig:schematics}
\end{figure*}

\section{Synthetic Stellar Spectra}
\label{sec:data}

Our training data consists of APOGEE-like synthetic high-resolution stellar ($\rm R \sim 22,000$) spectra generated using the Kurucz models \citep{Kurucz1981,Kurucz2005,Kurucz2017}. Kurucz models solve for a 1D stellar atmosphere with {\tt ATLAS2}. The atmosphere is then used to generate realistic spectra through a radiative transfer model with {\tt SYNTHE}, assuming Saha and Boltzmann equilibrium. The spectral simulator allows us to access a varied stellar feature space as input, namely in effective temperature, surface gravity, metallicity, and elemental abundance space consisting of C, N, O, Na, Mg, Al, Si, P, S, K, Ca, Ti, V, Cr, Mn, Fe, Co, Ni and Cu. We generate 20,000 spectra by uniformly sampling in temperature between 4300 and 4500~K, surface gravity between 1.9 and 2.2~dex, and elemental abundance between -0.1 and 0.05~dex for all elements apart from nitrogen, which we sample between 0.0 and 0.15~dex. The native resolution of the Kurucz models is of $R = \lambda/\Delta \lambda = 300,000$, where $\lambda$ is the wavelength. We convolve this model using the mean line spread function from the APOGEE observations \citep{Holtzman2015}. We further assume a Nyquist sampling; we store one pixel per APOGEE resolution element $\Delta \lambda$, leading to a spectrum with 2405 pixels.

To mimic actual observations, we degrade our spectra to a signal-to-noise ratio of 300, corresponding to the level of high-fidelity observed APOGEE spectra. Recall that our goal is to construct a conditional distribution of spectra given stellar parameters. Since the ground truth stellar parameters are often unknown in actual data (but can be estimated from the color-magnitude diagram or the spectra themselves), we also assume a 25~K scatter in effective temperature, 0.1~dex in surface gravity and 0.01~dex in metallicity.
\begin{figure*}[t]
    \centering
     \includegraphics[width=0.85\textwidth]{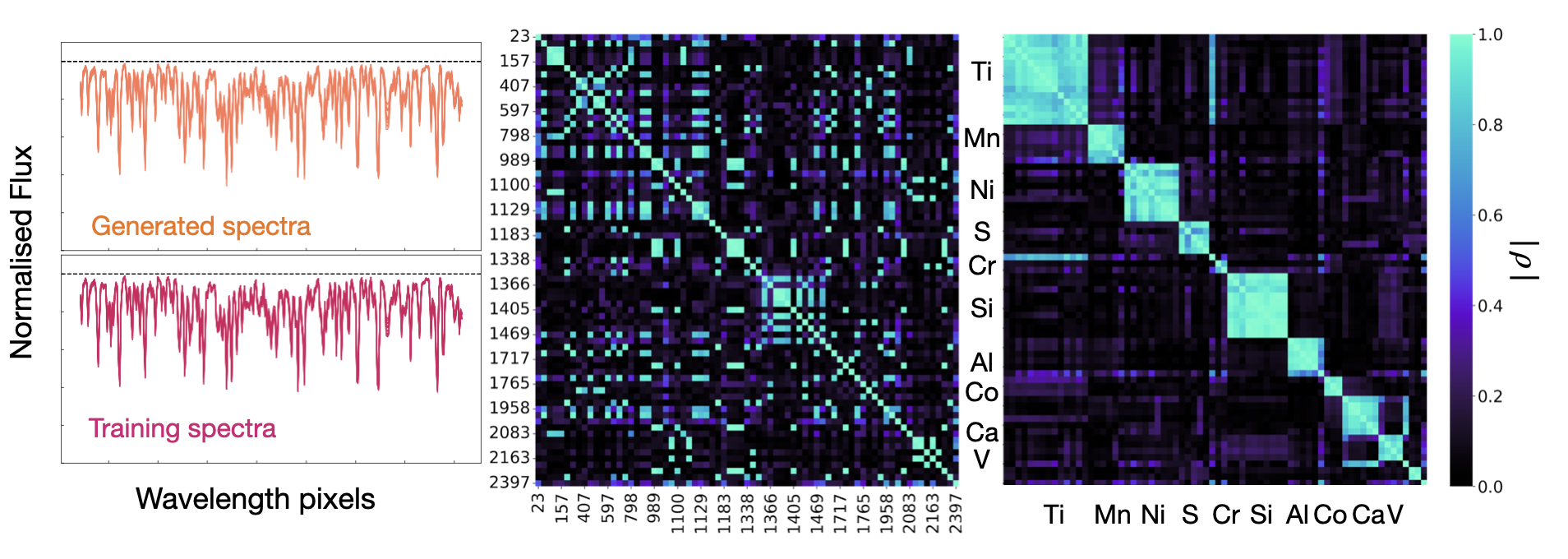}
     \vskip -0.1in
     \caption{Our generative models produce realistic stellar spectra (upper left panel) indistinguishable from the training spectrum (lower left panel). The probabilistic nature of our models further allows the calculation of the correlation structures between wavelengths and reveals which atomic transitions at different wavelengths are contributed by the same elements. The middle panel shows the correlation matrix sorted by wavelengths, and the right panel by elements. The re-indexing is done by taking the former as an adjacency matrix of a graph.}
     \vskip -0.1in
     \label{fig:model}
\end{figure*}

\section{Normalizing Flows for Stellar Spectra}

In this study, we assume a normalizing flow model to solve an unsupervised\footnote{Since this study also assumes stellar parameters, some might argue that the model should be regarded as weakly supervised. However, as the stellar parameters are only used as conditional variables, we opt to stick to the unsupervised terminology.} task of learning a model that best represents the distribution $p(\mathbf{x})$ from which the training samples $\{\mathbf{x_i}\}$ were drawn. At its core, the normalizing flow aims to find a change of variable $f_\theta$, such that the transformed variable $\mathbf{z} \equiv f_\theta(\mathbf{x})$ is normally distributed, i.e. $p_z(\mathbf{z}) = \mathcal{N}(\mathbf{O},\mathbf{I})$, effectively finding an invertible function that ``Gaussianize" the distribution \citep{Dinh2014, Dinh2016, Kingma2018}.

The training task of normalizing flows involves optimizing the parameters of $f_\theta$ by maximizing the log-likelihood over the training set $\mathcal{D}$,  $\mathcal{L}(\mathcal{D}) = \frac{1}{\mathcal{D}}\sum_{x \in \mathcal{D}}{\rm log} \ p(\bf x)$, where
\begin{equation}
  p_{\rm x}({\bf x}) = p_{\rm z}(f_\theta({\bf x}))|{\rm det} \ J_{f}({\bf z})|
  \label{eq:eq2}
\end{equation}
and $J$ the Jacobian of the transformation.

In our study, we have $\mathbf{x} = (\mathbf{S},\mathbf{y}) \in \mathbb{R}^{2405+3}$, where $\mathbf{S}$ is the stellar spectrum, and $\mathbf{y}$ three stellar parameters, namely the effective temperature, surface gravity and the metallicity. Learning the joint distribution $p(\mathbf{S},\mathbf{y})$ allows us to evaluate the conditional distribution $p(\mathbf{S}|\mathbf{y})$ which will prove critical. While in theory it is possible to model the conditional distribution directly as done, for example, in \citet{Ting2021}, we found that in practice training a conditional distribution on this large dimensionality is prohibitively difficult, which has motivated us to train on the joint distribution instead.

{\bf Architecture:} Extensive exploration on architectures such as RealNVP \citep{Dinh2016}, SurVAE \citep{Nielsen2020} and Neural Spline Flows \citep{Durkan2019} have led to our final architecture, which we dub the name \mendis{}. \mendis{} composes of an Activation Normalization (ActNorm) layer and a GLOW (`Conv1x1') operation, both introduced in \citet{Kingma2018}, followed by a Neural Spline Flow. The ActNorm layer performs an affine transformation and is used to improve convergence during the training task. GLOW is a block-diagonal linear transformation that essentially rearranges and mixes the dimensions of the input vector before reaching the coupling layer to maximize information sharing across all the input variables. Finally, the Neural Spline Flow employs a monotonic rational-quadratic transform that performs a cubic spline transformation on the input \citep{Durkan2019}. These transforms end up in a tractable Jacobian associated with their lower diagonal matrix and can be inverted in a single pass. Hence they allow for fast likelihood evaluation and sampling, which makes them a powerful choice.

{\bf Training approach:} A key advantage of normalizing flows, as opposed to other generative models such as Generative Adversarial Networks, is that the maximum likelihood loss function is resilient against mode collapse \citep{Bond-Taylor2021}. However, this comes with the cost. As already alluded to in Equation~\ref{eq:eq2}, normalizing flows require neural networks to be invertible and have tractable and efficiently computable Jacobian. This stringent constraint often makes normalizing flows less expressive. As such, we found that an extensive normalizing flow is needed to deal with the considerable dimensions in the spectral space where $x \in \mathbb{R}^{2405+3}$. We choose a normalizing flow with 20 units of ActNorm, GLOW and Neural Spline coupling layer. This architecture choice leads to a final network with approximately a billion parameters.

Our code is made multi-node multi-GPU parallelized across 16 Nvidia A100 GPUs using the Distributed Data-Parallel functionality of {\tt PyTorch} to train this massive network. We optimize our parameters using {\tt Adam} \citep{kingma_2014}, adopt a learning rate of $5 \times 10^{-6}$ and train for 3,000 epochs. Training across 16 A100 takes approximately 7 hours. All our codes are made available on Github.

\section{New Frontiers in Stellar Spectroscopy}
In the following, we present two applications to demonstrate the potential of normalising flows for stellar spectroscopy by harnessing their unique capabilities of describing the underlying probabilistic distribution of the spectra.

{\bf Unsupervised Learning of Atomic Features:} Although spectra are vectors with thousands of dimensions (pixels at different wavelengths), they are also remarkably simple. Most pixels are highly correlated due to the underlying physics: the absorption features are formed from different atomic transitions, resulting from quantum mechanics. As various elements have distinct atomic transitions at different wavelengths, the underlying correlation structures of spectral are therefore tell-tale signs of contributions from specific elements, but with one caveat. Apart from the elemental abundances, stars with different stellar parameters will also alter the opacity of the stellar photosphere and hence the emergent spectrum. Therefore, if we were to evaluate the correlation matrix from $p(\mathbf{S})$, the correlation matrix is dominated by the contribution of the stellar parameters, obscuring the contributions from individual elements.

Modeling the training spectra with \mendis{} enables a new window to tackle this problem. More specifically, once we have trained a robust statistical description of $p(\mathbf{S},\mathbf{y})$, the model allows us to evaluate all moments of the conditional distribution $p(\mathbf{S}|\mathbf{y})$ through repeated sampling from the normalizing flows at a fixed $\mathbf{y}$. The conditioning will allow us to eliminate all influence on the spectra from the stellar parameters, further revealing only contribution from the chemical composition of the stars. Importantly, this statement holds even if the unsupervised model itself is trained with a sample without any two stars sharing the same $\mathbf{y}$.

The left panels of Fig.~\ref{fig:model} demonstrates how well \mendis{} perform in terms of spectral reconstruction. The bottom panel shows ten examples of a training spectrum $\tilde{\mathbf{x}}$, and the top panel shows their closest counterparts determined by computing ${\rm argmin}_{\mathbf{z}}|\tilde{\mathbf{x}}-f_\theta^{-1}({\mathbf{z}})|^2$ from spectra $f_\theta^{-1}({\mathbf{z}})$ generated from \mendis{}. The two are indistinguishable from each other, demonstrating that our normalizing flow is a faithful surrogate of the training spectra.

The middle panel further demonstrates the correlation matrix evaluated from $p(\mathbf{S}|\mathbf{y})$, where we fix the conditioning stellar parameters $\mathbf{y}$ at an effective temperature of $4300$~K, surface gravity $2.1$~dex and solar metallicity. The middle panel shows the native correlation matrix sorted by the pixel wavelengths. We only show a subset of pixels with strong absoprtion features. The correlation matrix is sparse and unstructured because atomic transitions from the same elements can occur at vastly different wavelengths. To better visualize the high-correlation clusters, we then turn the correlation matrix into an adjacency matrix of a graph and assume an edge between the two pixels if their correlation is larger than 0.8. We then use the adjacency matrix to re-index the original correlation matrix, essentially searching for connected trees within the graph. The re-indexed version of the correlation matrix is shown on the right.

The right panel shows that, by fixing the stellar parameter, \mendis{} reveals the striking underlying correlation from the spectra. Different elements are sorted as block diagonals in the correlation matrix. Notably, the high correlation structure endures even though we have trained using noisy spectra and labels. The result demonstrates that unsupervised learning could unearth unknown atomic transitions through empirical data, even though we are entirely agnostic about their elemental abundances.\par
\begin{figure}
     \includegraphics[width=\columnwidth]{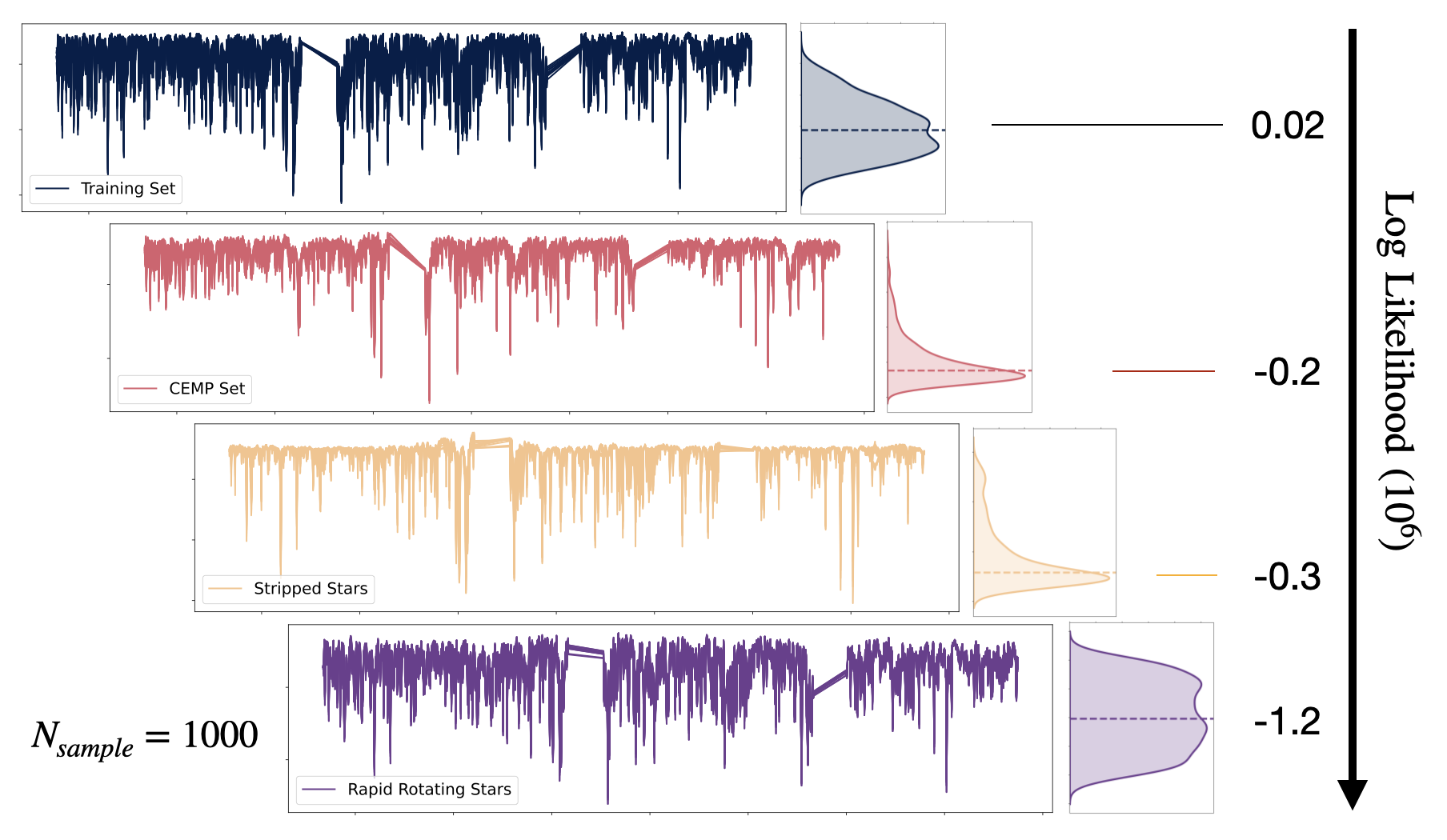}
     \vskip -0.1in
     \caption{Out-of-distribution detection with \mendis{}. Various outlier classes, such as CEMP stars, stripped stars and rapidly rotating stars are assigned to have a low likelihood by \mendis{} compared to the training data, even though their spectra, as shown in the inset plots, might not look too dissimilar. Without any predetermined stellar labels, \mendis{} provides an automated way to triage outlier objects of interest, sifting through large spectroscopic data.}
     \vskip -0.2in
     \label{fig:outlier}
\end{figure}

{\bf Out-of-distribution detection:} We have seen many extensive spectroscopic surveys come to fruition in recent years, collecting tens of millions of spectra. With this vast spectroscopic data, there are bound to be many unknown unknowns lurking in the data. However, supervised learning tends to project all data to the model known space, limiting our ability to find the exciting outliers.

\mendis{}, as an unsupervised method, provides a new opportunity to sift through high-dimensional information. It identifies unlabelled unique objects without requiring label determination beforehand. As \mendis{} describes the distribution of spectra $p(\mathbf{S})$, the likelihood for any object can be simply evaluated according to Equation~(\ref{eq:eq2}). Specifically, for any new datum $\mathbf{\tilde{S}}$, the likelihood $p(\mathbf{\tilde{S}})$ evaluates if $\mathbf{\tilde{S}}$ is within the distribution $p(\mathbf{S})$ spanned by the training set, or if it is something that the model has not encountered.

As a proof of concept, here we built three classes of some of the more sought after known unknowns in the study of galaxy evolution \citep[e.g.,]{Carollo2014,Gotberg2019}. Particularly, we generate from the Kurucz models spectra from carbon-enhanced metal-poor (CEMP) stars ($-2.0 < {\rm [Fe/H]} < -1.5$ and $1 < {\rm [C/Fe]} < 2$), rapid-rotating stars ($v_{\rm macro} > 20\,$km/s), and stripped stars (effective temperature$\,=4500-7000\,$K, surface gravity$\,=4.5 - 6$). Fig. \ref{fig:outlier} shows that directly using only the spectra, \mendis{} successfully discriminates between the training set and these extreme-valued test samples. \mendis{} assigns a low likelihood to these outlying test samples, signaling that they are objects of interest and merit follow-ups.

While we study known unknowns in this case study, we emphasize that \mendis{} is completely agnostic about stellar labels. It simply assigns a low likelihood to spectra that do not look like anything else in the training data and can be generalized to true unknown unknowns. Our results demonstrate that \mendis{} can serve as a generic spectral broker for any spectroscopic surveys, independent to existing label determination pipelines, to triage for outliers.

\section{Prospects and Future Directions}
A common criticism of unsupervised learning is that it often lacks interpretability. This study shows that this needs not to be the case. Unsupervised learning with normalizing flows retains its interpretability, as demonstrated by \mendis{}'s ability to identify pixel correlations. Besides, \mendis{} also provides a more principled way to perform out-of-distribution detections. But more generally, the statistical description of the distribution of spectra through \mendis{} has a wide range of applications beyond the few case studies explored here. For example, \mendis{} can serve as a bridge for supervised and unsupervised learning -- the learned distribution can serve as the prior distribution for domain adaption to close the model-data synthetic gap \citep{Obriain2021}. Additionally, the distribution can serve as the basis for semi-supervision and few-shots learning with a limited number of stars with high-fidelity labels.

Although \mendis{} holds many promises, it has much room for improvement. Notably, since spectra lack obvious physical symmetry, currently, a brute-force large normalizing flow with a billion parameters is chosen to depict the spectral distribution robustly. Recent normalizing flow attempts in cosmology have taught us that embedding physical knowledge (in cosmology, spherical symmetry) can significantly simplify the architecture \citep{Dai2022}. Since most information in spectra is encoded in their pixel correlations, we posit that preprocessing the data with self-attention modules to pre-learn the pixel correlations might lead to a more compact architecture. Despite all the challenges, deep normalizing flows pave the way to fully harnessing information from the ongoing and forthcoming spectroscopic surveys (4MOST, SDSS-V), allowing us to finally reach the stars.

\bibliography{example_paper}
\bibliographystyle{icml2022}
\end{document}